\documentclass[12pt]{article}
\usepackage{times}
\usepackage{geometry}  
\usepackage[utf8]{inputenc}
\usepackage{enumitem,amssymb}  
\usepackage{ragged2e,xcolor}
\usepackage{graphicx} 
\usepackage{upgreek}

\RequirePackage[sort&compress]{natbib}
\RequirePackage{natbib}

\newlist{thematic}{itemize}{8}
\setlist[thematic]{label=$\square$}
\usepackage{pifont}

\begin{document}

\newcommand{\CII}{C\,{\sc ii}}
\newcommand{\OI}{O\,{\sc i}}
\newcommand{\HII}{H\,{\sc ii}}

\raggedright
\huge
Astro2020 Science White Paper \linebreak

\color{blue}Stellar Feedback in the ISM Revealed by Wide-Field \mbox{Far-Infrared}  Spectral-Imaging\linebreak
\color{black}

\small

\noindent \textbf{Thematic Areas:} \hspace*{60pt} 
$\boxtimes$ Resolved Stellar Populations and their Environments \hspace*{40pt} 

  
\textbf{Principal Author:}

Name: \color{blue}{Javier R. Goicoechea}\color{black}	
 \linebreak						
Institution:  Instituto de F\'{\i}sica Fundamental, CSIC, Madrid, Spain.
Email: javier.r.goicoechea$@$csic.es
 
\textbf{Co-authors:} 
  \linebreak
\color{blue}{Maryvonne Gerin,} \color{black} Observatoire de Paris \& CNRS, France.\\
\color{blue}{Emeric Bron,} \color{black} Observatoire de Paris-Meudon, France.\\

\justify

  \textbf{Abstract:} The radiative and mechanical interaction of stars with their environment drives the evolution of the ISM and of galaxies as a whole.
The far-IR emission ($\lambda \simeq$30 to 350\,$\mu$m)  from atoms and molecules dominates the cooling of the warm gas in the neutral ISM,
 the material that ultimately forms stars.
\mbox{Far-IR} lines are thus the most sensitive probes of stellar \textit{feedback} processes, and allow us to quantify the deposition and cycling of energy in the ISM.
While \textit{ALMA} (in the (sub)mm) and \textit{JWST} (in the IR) provide
astonishing sub-arcsecond resolution images of point sources and their immediate environment, they cannot access the main interstellar gas coolants, nor are they designed to image entire star-forming regions (SFRs).  \mbox{\textit{Herschel}} 
far-IR photometric images of the  interstellar dust thermal  emission revealed the ubiquitous \mbox{large-scale} filamentary structure of SFRs, their mass content, and the location of thousands of prestellar cores and protostars. These images, however, provide  a \textit{static} view of the ISM: not only they dont  constrain the cloud dynamics, moreover they cannot reveal the chemical composition and energy transfer within the cloud, thus giving little insight into the regulation process of star formation by stellar feedback.
 In this \textit{WP} we emphasize the need of a space telescope with wide-field spectral-imaging capabilities in the critical far-IR domain.

\begin{figure*}[ht]
\centering
\includegraphics[scale=0.14, angle=0]{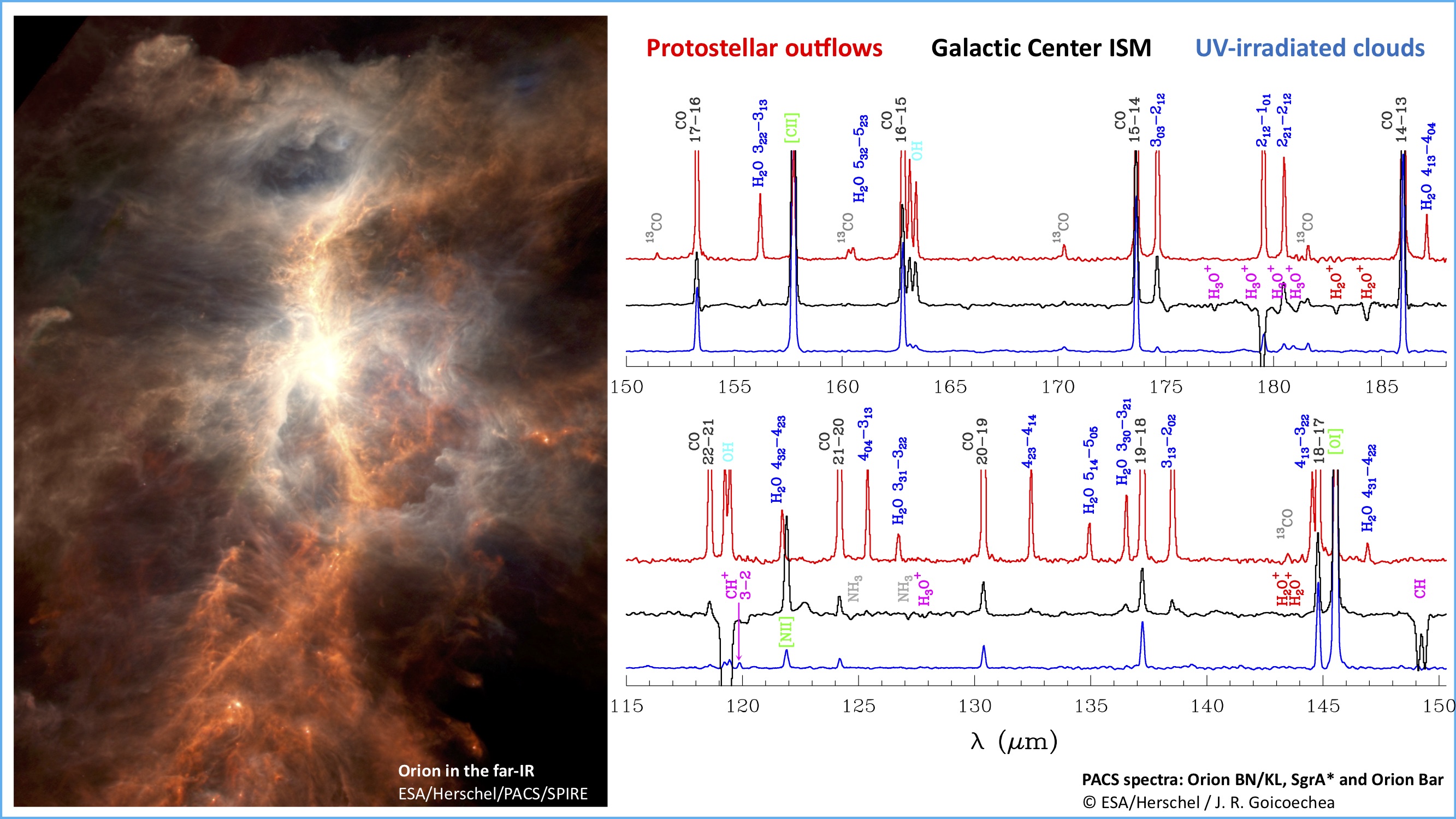}
 \label{fig:cover}
\end{figure*}

\pagebreak

\textbf{ \large 1. Far-IR Tracers of Stellar Feedback: Warm Molecular Gas}

\justify

\small

Massive stars dominate the injection of radiative and mechanical energy into the ISM through ionizing radiation, stellar winds, supernova explosions, and merger encounters \citep{Beuther07,Zinnecker07,Krumholz14,Motte18,Pabst19}. This energy input shakes the environment, heats the gas, disrupts star-forming regions (SFRs), and creates the cloud and intercloud phases of the ISM.  Massive stars are born inside  giant molecular cloud (GMC) cores. Protostars of different masses develop inside these star-forming cores \mbox{\citep[e.g.,][]{McKee07}}. Their  outflows shock the ambient cloud, heating and compressing the immediate environment  around them. This is 
an example of small spatial scale protostellar feedback.
Young stellar objects (YSOs) can be detected in photometric images by the bright far-IR/submm luminosity of their heavily obscured dusty cocoons. 

On the large-spatial scales of an entire GMC, however, most of the gas and dust  emission does not arise from individual YSOs but from the extended and more massive cloud component -- the poorly known star-forming \textit{environment}. 
 Once a new massive star or cluster is formed, the energy and momentum injected by UV photoionization, radiation pressure, and stellar winds, ionize and erode the parental molecular cloud, creating \HII~regions and blowing expanding bubbles \citep[see  simulations in e.g.,][]{Krumholz14,Rahner17}. \mbox{Stellar} feedback 
can thus determine the gas physical conditions and chemical composition
 at large scales, drive  the evolution of the parental cloud itself, and regulate the formation of new stars. However, due to the lack of wide-field spectroscopic observations, there are fundamental aspects that remain to be understood. UV radiation in particular, is a very important player in the interaction of stars and ISM.

 \mbox{Photodisociation} regions (PDRs)  develop at the interfaces between the hot ionized gas and the cold molecular gas, and more generally, at any slab of neutral gas (\mbox{hydrogen} atoms not ionized) illuminated by stellar far-UV (FUV) photons  
with \mbox{$E< 13.6$\,eV} \citep[][]{Hollenbach97}. 
The~famous far-IR [\CII]158\,$\upmu$m fine-structure line  is typically the brightest emission  from PDR gas, and it dominates the cooling of the neutral ISM at the scales of an entire galaxy. 
It thus provides key information on the energy deposited in the ISM \citep[see  \textit{COBE}'s low resolution maps of the Milky Way,][]{Bennett94}. With an ionization potential of 11.3\,eV,  however, C$^+$ can also be abundant in both the warm neutral or ionized gas  \mbox{\citep[e.g.,][]{Pineda13}}. Hence,  it is not always trivial to delimitate the origin
of the [\CII]\,158\,$\upmu$m emission  and fully exploit its diagnostic power. 
Because of their specific chemistry, \textit{certain molecular species that
we discuss below, emitting in the far-IR,  are very good \textit{filters} of the different feedback processes}.

\textit{\textbf{Finding observational tracers of the feedback processes acting on the molecular gas, the fuel to form new stars, is crucial not only locally, but also in the framework of star formation across cosmic time}}.\\

\textbf{Only in the far-IR and only from space.} While the low-energy molecular lines that can be detected from (sub)mm-wave ground-based radio telescopes such as ALMA typically trace cold and quiescent interstelar gas, the extended \textit{\textbf{warm molecular gas}} (\mbox{$T_{\rm k}\gtrsim 100$\,K}) affected by feedback processes (i.e., heated by stellar UV fields, enhanced \mbox{X-ray} doses, cosmic-ray particles, or affected by shocks and turbulence dissipation) naturally emits at higher frequencies, in the far-IR. The far-IR domain  hosts  a  plethora of bright atomic fine-structure lines (from the neutral and ionized phases of the ISM) and high-energy (rotationally-excited)  lines from CO, H$_2$O, CH$^+$, HD, and other hydrides that cannot be detected from the ground. These \mbox{far-IR} lines, often the most luminous lines emitted by the ISM of galaxies as a whole, prove to be unique diagnostics of the different types of energy and momentum input deposited into the ISM. That is, of the different feedback mechanisms.
The launch of ESA's \textit{Infrared Space Observatory} (\textit{ISO}) in 1995 opened the complete far-IR domain  to spectroscopic observations, and demonstrated that
the gas properties and energetics of YSOs can be constrained by studying their \mbox{far-IR} rotationally excited  CO and H$_2$O line emission
\mbox{\citep[e.g.,][]{Ceccarelli96,Giannini01}}. These lines are major coolants of the $hot$ ($T_{\rm k}$$>$500~K) molecular gas and the unambiguous  signature of shocked gas  \mbox{\citep[e.g.,][]{Kaufman96}}. \mbox{IR H$_2$} lines, ~observed by \textit{ISO},  NASA's \textit{Spitzer} and soon again with \textit{JWST}, are also excellent tracers of the hot molecular gas, but their emission is often heavily affected by dust extinction toward embedded protostars. \textit{ISO} had a primary mirror of $\sim$60\,cm size and carried primitive far-IR detectors. This limited the angular resolution and sensitivity of these pioneering detections.
The launch of ESA's \textit{Herschel} in 2009, with a
much  larger $\sim$3.5m telescope and equipped with more sophisticated far-IR instrumentation (direct detection detectors and heterodyne receivers) allowed a more robust characterization of (a rather short list of) low-and high-mass YSOs in different stages of evolution \mbox{\citep[e.g.,][]{vanDishoeck11}}. However, robust statistics and  spectroscopic observations of much larger protostellar samples 
and their environment  are largely missing. More precisely,  \textit{\textbf{{Herschel} did not have instrumentation  to  carry out  spectroscopic maps of entire star-forming regions}}. 
 
 \begin{figure*}[ht]
\centering
\includegraphics[scale=0.155, angle=0]{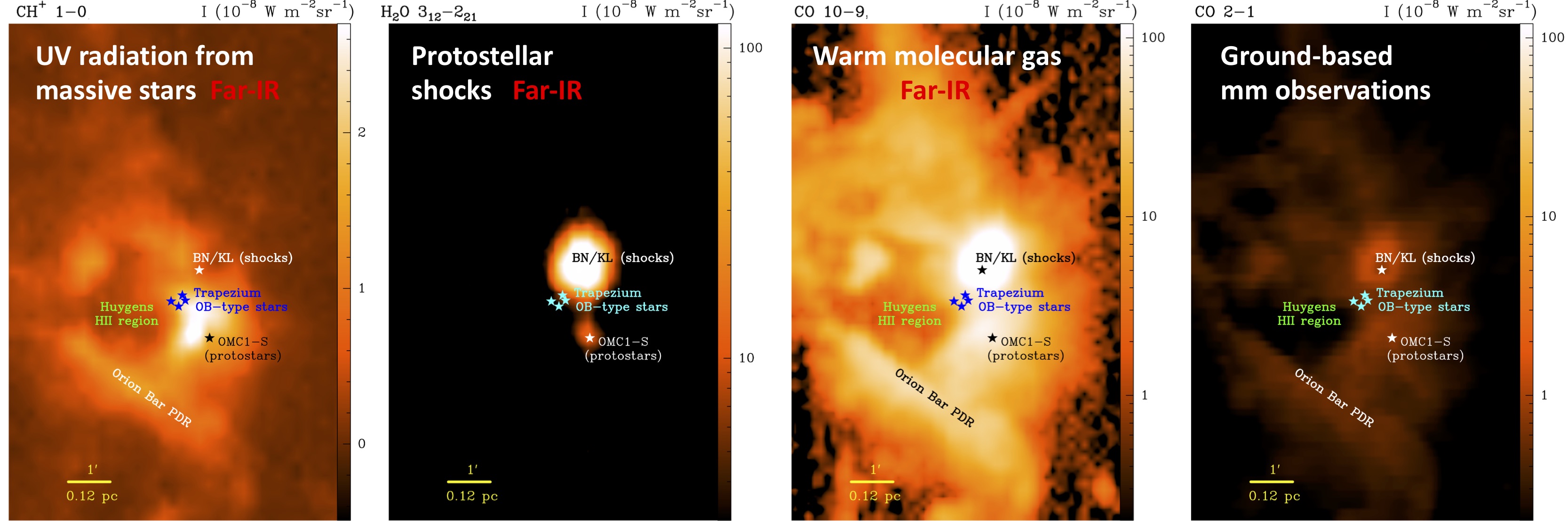}
 \caption{\small \textit{Herschel} images of different molecular lines toward Orion (the closest
 high-mass star-forming region). They all show different spatial distributions: 
CH$^+$ (UV-irradiated  gas), H$_2$O (hot shocked gas), \mbox{CO $J$\,$=$\,10--9} (extended warm gas) and CO $J$\,$=$\,2--1 (observed from the ground and much fainter).
}\label{fig:OMC1-maps}
\end{figure*}

Figure~\ref{fig:OMC1-maps} shows $\sim$85~arcmin$^2$ images of  key far-IR line  obtained with the HIFI receiver toward the Orion molecular cloud core \mbox{(OMC-1)}, the closest high-mass SFR (at  $\sim$414\,pc). These are one of the few spectroscopic-maps obtained by
\textit{Herschel} \citep{Goi19}.
These images show the emission from CH$^+$
(strongly FUV-irradiated gas),  H$_2$O (hot shocked gas from protostellar outflows), CO~$J$\,$=$\,10--9 \mbox{(extended warm PDR gas)} and CO~$J$\,$=$\,2--1 (observed from the  ground). 
  The four emission lines show remarkably different spatial distributions, \textit{\textbf{emphazising the distinctive diagnostic power of different molecules and far-IR lines}}.  
In the following, we briefly review the information that can be obtained \textit{only} from far-IR line observations, and we apply it  to our case:
the  radiative feedback from massive stars. \vspace{+0.1cm}\\
\textbf{H$_2$O and OH} only reach high abundances in hot/warm shocked gas.  \mbox{Far-IR} H$_2$O and OH rotational lines cover a  broad range of energies and excitation conditions. Thus, they are excellent diagnostics of YSOs, and of shocks due to outflows in general:
protostellar, supernovae or extragalactic \citep{Sturm11}. Owing to the water vapor on Earth's atmosphere, these lines can not be observed from the ground or with \textit{SOFIA}.\vspace{+0.1cm}\\
\textbf{CO rotational ladder} can be used as a ``thermometer" of the hot/ warm molecular gas. Far-IR \mbox{high-$J$~CO} lines have been detected 
by \textit{ISO} and \textit{Herschel} at different spatial scales: from outflows driven by nearby low-mass protostars to luminous active galactic nuclei (AGN) galaxies. In these energetic sources affected by shocks, the CO spectral line energy distribution (SLED) peaks at $J>15$  
($\lambda <$173.6~$\upmu$m) and shows detectable emission at 
 $J>30$ \citep[$\lambda <$87.2~$\upmu$m,][]{Hailey12,Karska13}.
The CO line emission measured by \textit{Herschel} toward less extreme SFRs, however, typically peaks at $J\lesssim10$  and shows  large-scale emission as seen in nearby SFRs like Orion (see Fig.~\ref{fig:OMC1-maps}).
When observed at high spectral resolution ($<$1~km\,s$^{-1}$), 
\mbox{mid-$J$ CO} lines show narrow line-profiles, demonstrating that they arise
from FUV-illuminated extended warm gas and not from fast shocks \cite[e.g.,][]{Goi19}.\vspace{+0.1cm}\\
 \textbf{Hydrides such as CH$^+$} have their rotational lines
 in the far-IR.
 These molecules are not only important because their formation represents the first steps of interstellar chemistry \mbox{\citep[see][]{Gerin16}}.  \textit{Herschel} showed that their abundances are also unique tracers of the  H$_2$ column density   (HF and CH), the ionization rate by cosmic-ray particles or \mbox{X-rays} (OH$^+$, H$_2$O$^+$ or ArH$^+$), and the stellar FUV  field (CH$^+$ or SH$^+$). 
 In addition, the hydride emission from dense star-forming clouds
($n_{\rm H}$\,$>$\,1000~cm$^{-3}$) traces \textit{\textbf{gas components
and physical conditions that cannot be studied from the ground}}. One outstanding example is CH$^+$.
 The main gas-phase pathway  producing detectable quantities of CH$^+$  is reaction \mbox{C$^+$~+~H$_2$\,($v$)~$\rightarrow$~CH$^+$~+~H},
where $v$ refers to the specific H$_2$ vibrational level  \citep{Sternberg_1995,Agundez10}. 
For H$_2$ molecules in \mbox{$v$}=0 state, this reaction is very endothermic,
\mbox{$\Delta E/k$\,$\simeq$\,4300~K}, much higher than the typical gas temperatures of molecular clouds.  However, the reaction becomes exothermic and fast for $v$\,$\geq$\,1.
Observations and theory show that UV photons from nearby massive stars can radiatively pump H$_2$ to vibrationally excited states $v$\,$\geq$\,1 over large spatial-scales. In consequence, CH$^+$ can be abundant at large scales too,
and its emission can be used to characterize the  stellar UV radiation field
\mbox{\citep{Goi19}}. In particular, CH$^+$ probes a narrow (low A$_V$) but very extended component of UV-irradiated molecular cloud surfaces (where not all carbon is  locked in CO). CH$^+$ is clearly a unique tracer of harsh
interstellar conditions. Not surprisingly, CH$^+$~($J$=1--0) emission has been 
detected by ALMA toward high-redshift ULIRG galaxies \mbox{\citep{Falgarone17}}. Hence, CH$^+$ can also be used to constrain the energetics of the primitive ISM. \vspace{0.2cm}\\
\textbf{Radiative Feedback and PDRs}: Most of the mass contained in GMCs resides at low visual extinction depths (A$_V$). Therefore, most of the gas is permeated by stellar FUV photons.
A strong FUV radiation field from nearby massive stars  induces a plethora of poorly understood dynamical effects and chemical changes in the cloud. 
The high thermal presures inferred from \textit{Herschel} observations toward a few \HII\,/PDR interfaces,
\mbox{$P_{\rm th}=T_k \cdot n_{\rm H} \gtrsim 10^7-10^9$~cm$^{-3}$\,K},
are consistent with the expected dynamical response of molecular clouds to strong \mbox{FUV} radiation: the cloud edges are heated and compressed, and
photoevaporate if the high pressures are not balanced by those of the  environment \citep[][]{Bertoldi96,Bron18}. 
Cloud photoevaporation  models predict that the  thermal pressure at the irradiated cloud edges  scales with the strength of the \mbox{FUV} flux  impinging the cloud (see Fig.~2). CH$^+$, a very reactive and short-lived molecule, proves to be as a unique tracer of these narrow  layers, 
clearly revealing feedback processes. In particular, \citet{Goi19} found a spatial correlation between the intensity of the CH$^+$ emission and $G_0$, the flux of FUV photons  (Fig.~2). The observed correlation is supported by PDR   models in the parameter space \mbox{$P_{\rm th}/G_0$\,$\approx$\,[5$\cdot$10$^3$--8$\cdot$10$^4$]~cm$^{-3}$\,K} where many  observed PDRs lie \citep[][]{Joblin18}.

\vspace{-0.4cm}
\begin{figure*}[h]
\centering
\includegraphics[scale=0.51, angle=0]{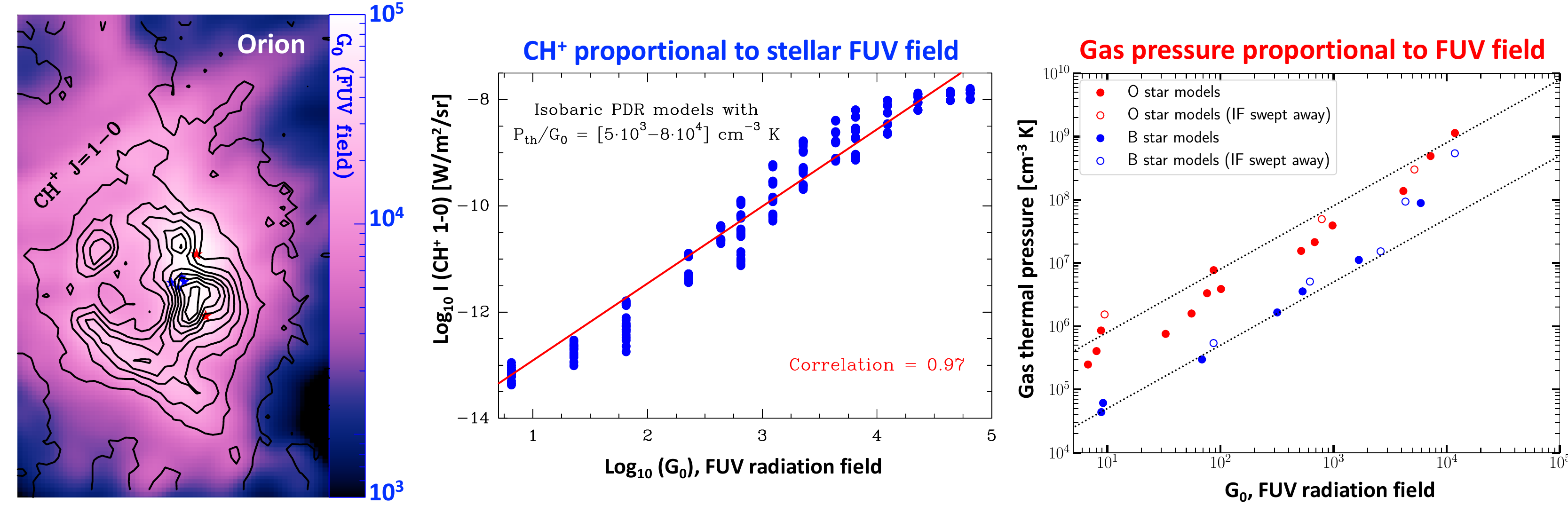}
\vspace{-0.5cm}
\caption{\small \textit{Left}:  
 FUV radiation field ($G_0$) arising from OB-type stars in Orion's Trapezium cluster. Black contours are the CH$^+$~($J$\,$=$\,1--0) emission
 observed by \textit{Herschel} and showing a very good spatial correlation with $G_0$. \textit{Middle}: This correlation is supported by PDR models \citep{Goi19}.  \textit{Right}: $P_{\rm th}$ vs. $ G_0$ correlation 
 at the edge of FUV-illuminated clouds
 predicted by  photoevaporation models  \citep{Bron18}.} \label{fig:Fig2}
\end{figure*}

\textbf{\large 2. Need and feasibility of far-IR spectral-imaging capabilities}\\ 
\small

The ISM is a central component of galaxies;  it provides the fuel to form new stars and it keeps a relic of previous star generations in the form of metal enrichment. 
Dust extinction hides  the star and planet formation processes from UV,  visible-light, and often near-IR observations.  
Stellar UV radiation, winds and supernova explosions heat and disrupt  the interstellar environment, sometimes enhancing, sometimes quenching the star formation processes. The main gas coolants  
in these  environments emit in the far-IR \citep[e.g.,][]{Schneider18}.
The peak of the dust emission also lies in the far-IR ($\lambda \simeq$100\,$\upmu$m for $T_{\rm d}$$\simeq$30~K). Therefore, 
\textit{\textbf{the far-IR is the natural domain to study the ISM  lifecycle and stellar feedback. Since these proceses take place at large-spatial scales, spectral-imaging capabilities are desperately needed}}. \textit{\mbox{Single-dish} space telescopes are the obvious choice for mapping and accessing the complete wavelength domain at high sensitivity and stability}.
The spectacular large-scale photometric-images of the dust emission taken by \mbox{\textit{Herschel}'s} far-IR cameras provided a static ``snapshot'' of the star-formation processes. 
However, these images say less about the stellar feedback processes  in the ISM, and how to quantify them.  It is only by carrying out large spatial scale far-IR observations of the critical gas cooling lines, and of other key astrochemical probes such as CH$^+$, that we will be able to quantify these proceses and their energetics. 
In addition, by using
\mbox{sub-km\,s$^{-1}$} velocity-resolution spectrometers similar to
\textit{Herschel}/HIFI or \textit{SOFIA}/GREAT receivers, we will also be able constrain the gas kinematics and turbulence properties \mbox{\citep[see e.g.,][]{Goi19,Pabst19}}.\vspace{+0.2cm}\\
\textbf{The future.} Neither \textit{ISO} nor \textit{Herschel} could carry out wide-field spectral-imaging observations. The new generation multi-beam  instruments on board \textit{SOFIA}  or balloon experiments,  combined with efficient \mbox{on-the-fly} (OTF) mapping techniques, will allow  faster and
larger maps. 
However, these stratospheric telescopes do not cover the entire far-IR domain (e.g., they can't observe H$_2$O or CH$^+$), they are
severely limited by the available flying time (e.g., they can't map the 
Galactic plane in reasonable times),  and they are restricted to the detection of a few bright lines. To overcome these issues, several far-IR space-telescope concepts have been conceived to probe the ISM lifecycle and stellar feedback, some adapted to the bright and extended emission of the Milky Way and nearby galaxies, others pushing beyond to characterize the ISM of distant and primitive galaxies  \citep[see e.g., the EU far-IR Space Roadmap,][]{Rigopoulou17}:\vspace{+0.1cm}\\
(1.) A $\sim$1\,m-class space telescope equipped with multi-beam heterodyne receivers (thus providing \mbox{sub-km\,s$^{-1}$} spectral resolution) and employing OTF techniques to scan very large areas of the Milky Way's galactic plane and provide velocity-resolved images of the interstellar gas in
the brightest gas cooling lines simultaneously
\citep[see e.g.,][for the \textit{FIRSPEX} concept]{Rigopoulou16}.\vspace{+0.1cm}\\
(2.) A cooled $\sim$2.5\,m-class space telescope equipped with a very sensitive
grating spectrometer, accessing the entire far-IR band  at medium spectral resolution, but providing very little spectral multiplexing capabilities \citep[see e.g.,][for the \textit{SPICA} concept]{Roelfsema18}.\vspace{+0.1cm}\\
(3.) A cooled $\sim$6\,m-class space telescope equipped with ultra-sensitive
 grating spectrometers accessing the entire far-IR band for wide-field spectral-imaging at low-spectral resolution ($R$ of a few hundred) or at higher $R$
 for single beam  observations \citep[see e.g.,][for the \textit{OST} concept]{Battersby18}.\vspace{+0.15cm}\\
Concept (1.) is specifically designed for mapping the galactic ISM, and is only limited by sensitivity and available bandwidth to simultaneously map $N$ lines. Concept (3.), if equipped with  multi-beam heterodyne receivers, will fullfil the science cases presented in this WP.  
Concept (2.) or (3. without receivers) will be able to detect all relevant
far-IR gas cooling lines, retrieving their line intensities (thus accessing the gas energetics) but not the gas kinematics or line-profile information. 
Even at low to medium $R$, they will detect the far-IR atomic and molecular lines from the  warm interstellar gas  and  from hundreds of protostars, and propoplanetary disks. This has been demonstrated by previous observations toward Orion with the grating spectrometers \textit{ISO}/LWS 
($R \simeq 200$) and \textit{Herschel}/PACS  ($R \simeq 2000$).  
Future instruments of this kind, however, will have to develop  spectral-imaging techniques as efficient as possible (e.g., using steering mirrors).\\

\textbf{Science case:} With the exception of a few far-IR spectral-maps  toward nearby and bright high-mass SFRs, almost all square-degree areas of the Milky Way covered by \textit{Herschel} far-IR dust photometric images  do not have spectroscopic counterpart in the main gas cooling lines ([\CII]\,158\,$\upmu$m, [\OI]\,63\,$\upmu$m) and in critical molecular line tracers of stellar feedback (high-$J$ CO lines, H$_2$O, CH$^+$...). In order to fill this gap, we propose 
a program to carry out spectral-imaging of the most relevant high-mass SFRs of the Galaxy (typically located a kpc distances) and also to map young  superclusters in nearby galaxies  \mbox{(e.g., 30 Dor in the LMC)}. As an example, here we focus on the  
CO $J$=12--11 ($E_u/k =$431~K) line at $\lambda$=217\,$\upmu$m. This is the first CO rotational line that could be detected by a instrument such as \textit{SPICA}/SAFARI, but we note that the highest-energy lines at \mbox{$J>50$} would be detectable too. In addition to the many CO and H$_2$O lines available in the \mbox{far-IR band}, other key target lines  are
CH$^+$~$J$=1-0, 2-1, 3-2 and 4-3 (at $\lambda$=359, 179.6, 119.8, and 90.0\,$\upmu$m respectively) and HD~$J$=1-0 and 2-1 
(at $\lambda$=112.0 and 56.2\,$\upmu$m respectively).\vspace{0.2cm}\\
In this example, we require to map a $\sim$75\,pc$^2$ area of 
several template GMCs. This would imply scanning angular sizes
of 50$'$$\times$50$'$ for massive SFRs like W51 (at a distance of 5~kpc), but  sizes of 5$'$$\times$5$'$ toward GMCs of the LMC. 
\textit{Herschel} observations show that the \textit{extended} warm molecular gas emission
in Orion (at only 0.5~kpc) produces surface brightness of $\sim$10$^{-8}$~W\,m$^{-2}$\,sr$^{-1}$ in the mid-$J$~CO lines 
(see Fig.~1).  Assuming extended emission filling the beam of the different
types of telescope concepts ($\sim$1, $\sim$2.5, and $\sim$6\,m telescope sizes), we compute the line flux sensitivity needed to detect
this emission level toward SFRs of increasing distance
(of course some regions will have a brighter or a fainter line emission level  depending on the particular excitation conditions
and dominant feedback proceses).
Expectations are shown in \mbox{Figure~3}. Just for reference, we note that the line sensitivity of
\textit{Herschel}/PACS was several times \mbox{10$^{-18}$~W\,m$^{-2}$} (5$\sigma$/1\,hr). We see that in order to detect the warm molecular gas
emission beyond the Galactic Center with a  $\sim$2.5\,m telescope, we require line sensitivities of several \mbox{10$^{-19}$~W\,m$^{-2}$}, and an order of magnitude better to image the molecular ISM of nearby galaxies.
While the smaller telescope concept will be more flexible to map degree-size areas of the Milky Way at parsec-scale spatial resolution, a $\sim$6\,m telescope improves by a factor of $\sim$2 the angular resolution provided
by \textit{SOFIA} or \textit{SPICA}. This will allow us reaching,  for the first time in far-IR spectral line maps,  a $\sim$0.1\,pc ($\sim$20,000~AU) spatial resolution for all SFRs at distances below 2.5\,kpc. This scale has been
suggested to be the universal width of the filaments that build up the structure of star-forming  clouds \citep[e.g.,][]{Arzoumanian11}. 
In addition, such kind of telescope will allow us to spatially resolve the emission from hundreds of protostars and their outflows.

\begin{figure*}[h]
\centering
\includegraphics[scale=0.53, angle=0]{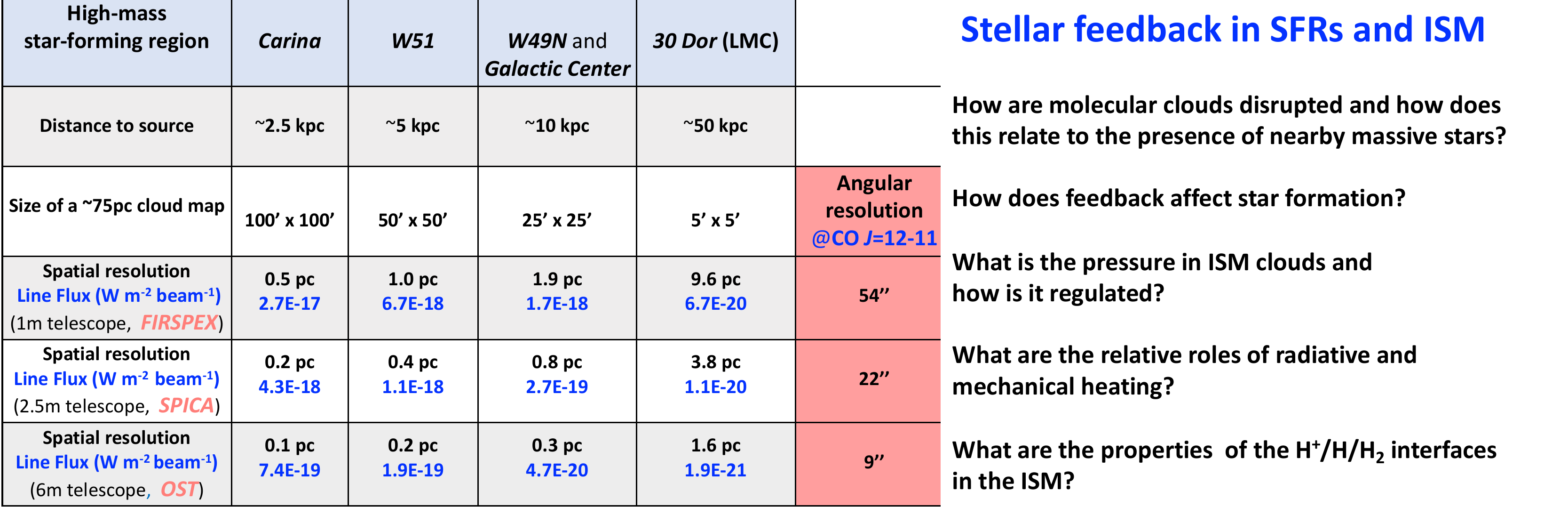}
\caption{\small Wide-field far-IR spectral-imaging of template high-mass star-forming regions. 
We propose to map a $\sim$75\,pc--size SFR
with three far-IR space telescope concepts. The resolution and required
line sensitivity are based on the CO $J$=12-11 line at 
$\lambda$$\simeq$217\,$\upmu$m
and its observed intensity toward Orion.
}\label{fig:Table}
\end{figure*}

\clearpage

\pagebreak

\bibliographystyle{aa}
\footnotesize
\bibliography{references}

\end{document}